\documentclass[10pt,aps,prl,twocolumn]{revtex4-1}

\usepackage{amsfonts}
\usepackage{graphics,graphicx, array, afterpage}
\usepackage{qcircuit,amsmath}
\usepackage{grffile}
\usepackage[export]{adjustbox}
\usepackage{sidecap}
\usepackage{braket}
\usepackage{multirow}

\begin{document}

\title{Comparison of Cloud-Based Ion Trap and \\ Superconducting Quantum Computer Architectures}
\author{S. Blinov}
\email{sdblinov@yahoo.com}
\author{B. Wu}
\email{bwu21@concordcarlisle.org}
\author{C. Monroe}
\email{c.monroe@duke.edu}
\affiliation{Duke Quantum Center, Departments of Electrical and Computer Engineering and Physics, Duke University, Durham NC 27701}

\date{\today}

\begin{abstract} 
Quantum computing represents a radical departure from conventional approaches to information processing, offering the potential for solving problems that can never be approached classically.  While large scale quantum computer hardware is still in development, several quantum computing systems have recently become available as commercial cloud services. We compare the performance of these systems on several simple quantum circuits and algorithms, and examine component performance in the context of each system's architecture.
\end{abstract}

\maketitle

\section{Introduction}
Quantum computing is a revolutionary form of information processing that is capable of solving some computational problems faster than conventional (classical) approaches \cite{MikeAndIke, Shor:1997}. Quantum information is represented by qubits, which can exist in superpositions of 0 and 1.  
Multiple qubits can be prepared in entangled states that generally possess an exponential number of superposed states, providing a quantum computer its power.
Quantum algorithms can be expressed in terms of circuits involving universal discrete quantum gate operations that entangle qubits, akin to wiring transistors together to perform logic operations in classical computers. 

Recently, gate-based quantum computers have become available as cloud computing services, hosted by IBM \cite{IBMqe} and Amazon Web Services \cite{AWS}. Two significantly different hardware types, ion traps and superconducting qubits, are featured on these cloud providers. Control over quantum circuits in each system is limited to certain quantum logic gate operations, and their qubits have significantly different architectures and error processes.

Here, we report the execution of several small quantum circuits and based on their performance, we investigate fundamental properties of these quantum systems such as qubit connectivity, gate noise and its accumulation, and State Preparation and Measurement (SPAM) error. 
Qubit noise, gate noise, and their accumulation is tested by performing increasing numbers of operations between two qubits. 
SPAM error is extracted by measuring qubits immediately after initialization in definite qubit states. 
Qubit connectivity is indirectly probed by implementing the Bernstein-Vazirani quantum algorithm \cite{BV} with various levels of circuit complexity.

\section{Quantum Computer Systems and Operations}
The physical architectures of these cloud based quantum computers fall into two categories: ion traps and superconductors. Both systems store information in a collection of qubits, with the two states $\ket{0}$ and $\ket{1}$ representing quantum states native to their particular hardware. The systems can express any quantum circuit in terms of a series of single- and two-qubit gate operations \cite{DiVincenzo:1995}. Our analysis focuses on the performance of two-qubit operations, as these typically dominate errors and facilitate communication between qubits.

\begin{figure}[b]
\includegraphics[width=\linewidth]{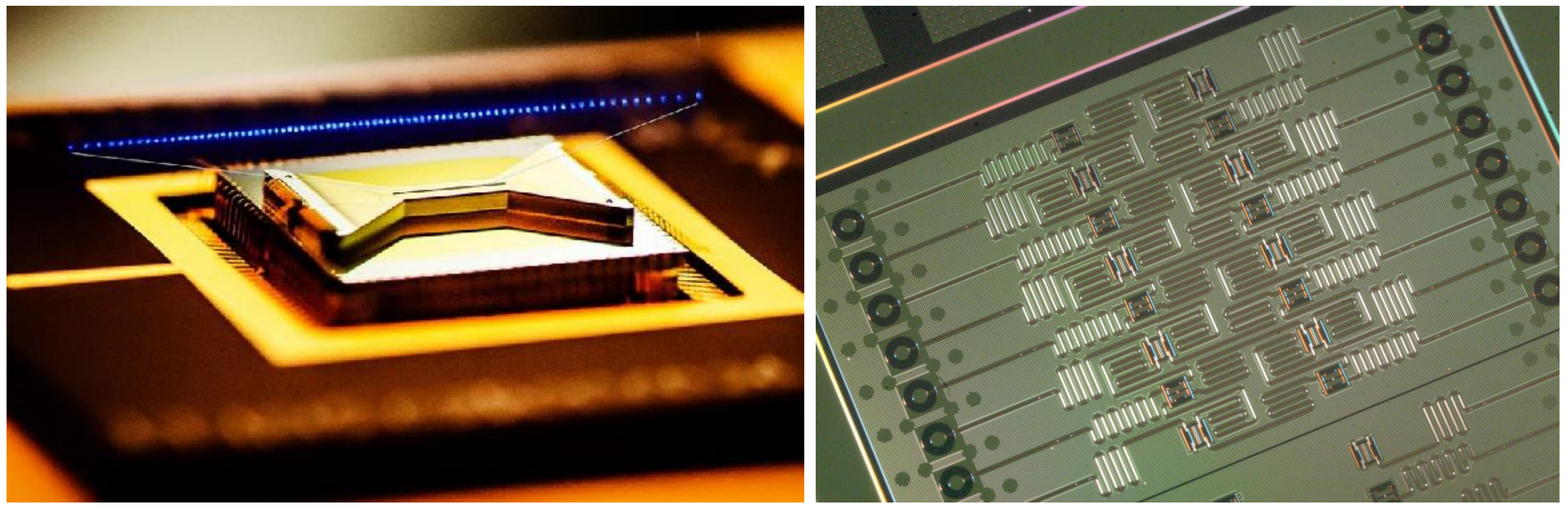}
\caption{Quantum computer cores available on the cloud. Left: Ion trap quantum computer chip from IonQ. Qubits are stored in electronic states of individual atoms (ions), shown glowing when laser light is applied. The ions are electromagnetically confined above a silicon chip (fabricated at Sandia National Laboratories), and qubit initialization, gates, and readout are realized with laser beams.  Right: Sample superconducting quantum computer chip from IBM. Qubits are represented by individual superconducting circuits, connected with electrical wiring on a 2D lattice. (Photos courtesy of IonQ and IBM.) \label{fig:QCs}}
\end{figure}

\begin{figure}[h]

  \begin{minipage}[b]{0.49\linewidth}
  \begin{eqnarray*} \\ 
    \ket{0} \hspace{-1pt}&\rightarrow&\hspace{-1pt} \cos\frac{\theta}{2}\ket{0}-ie^{-i\phi}\sin\frac{\theta}{2}\ket{1} \\ 
    \ket{1} \hspace{-1pt}&\rightarrow&\hspace{-1pt} \cos\frac{\theta}{2}\ket{1}-ie^{+i\phi}\sin\frac{\theta}{2}\ket{0} \\
  \end{eqnarray*}
   \vspace{\baselineskip}
  \mbox{\small \Qcircuit @C=2em @R=1.4em {\lstick{}& \gate{R(\theta,\phi)} & \rstick{}\qw}} \\
  \vspace{0.5\baselineskip}   (a)
  \end{minipage}
  \begin{minipage}[b]{0.49\linewidth}
  \begin{eqnarray*} \\ 
   \ket{00} \hspace{1pt}&\rightarrow& \hspace{8pt}\ket{00} \\ 
   \ket{01} \hspace{1pt}&\rightarrow& \hspace{8pt}\ket{01} \\
   \ket{10} \hspace{1pt}&\rightarrow& \hspace{8pt}\ket{10} \\
   \ket{11} \hspace{1pt}&\rightarrow& \hspace{-1pt}-\ket{11}
  \end{eqnarray*}
  \mbox{\small \Qcircuit @C=2em @R=1.4em {
  \lstick{}& \ctrl{1} & \rstick{}\qw \\
  \lstick{}& \gate{Z} & \rstick{}\qw
  }} \\
  \vspace{0.5\baselineskip}   (b)
  \end{minipage}
  \begin{minipage}[b]{0.49\linewidth}
 \begin{align} 
    \ket{00}  &\rightarrow& \hspace{-5pt}\cos{\chi}\ket{00} \hspace{-2pt}-\hspace{-2pt} i\sin{\chi}\ket{11} \nonumber\\ 
    \ket{01}  &\rightarrow&\hspace{-5pt} \cos{\chi}\ket{01} \hspace{-2pt}-\hspace{-2pt} i\sin{\chi}\ket{10} \nonumber\\
    \ket{10}  &\rightarrow&\hspace{-5pt} \cos{\chi}\ket{10}\hspace{-2pt} -\hspace{-2pt} i\sin{\chi}\ket{01} \nonumber\\
    \ket{11}  &\rightarrow&\hspace{-5pt} \cos{\chi}\ket{11}\hspace{-2pt} -\hspace{-2pt} i\sin{\chi}\ket{00} \nonumber
  \end{align}
   \vspace{\baselineskip}
  \mbox{\small \Qcircuit @C=2em @R=1.4em {
  \lstick{}& \multigate{1}{XX(\chi)} & \rstick{}\qw \\
  \lstick{}& \ghost{XX(\chi)} & \rstick{}\qw
  }} \\
  (c)
  \end{minipage}
  \begin{minipage}[b]{0.49\linewidth}
  \begin{align}  
    \ket{00} &\rightarrow&\hspace{-5pt} \cos{\chi}\ket{00}\hspace{-2pt} -\hspace{-2pt} i\sin{\chi}\ket{01} \nonumber\\     
    \ket{01}  &\rightarrow&\hspace{-5pt} \cos{\chi}\ket{01}\hspace{-2pt} -\hspace{-2pt} i\sin{\chi}\ket{00} \nonumber\\
    \ket{10} &\rightarrow& \hspace{-5pt}\cos{\chi}\ket{10}\hspace{-2pt} +\hspace{-2pt} i\sin{\chi}\ket{11} \nonumber\\
    \ket{11} &\rightarrow&\hspace{-5pt} \cos{\chi}\ket{11}\hspace{-2pt} +\hspace{-2pt} i\sin{\chi}\ket{10} \nonumber 
  \end{align}
   \vspace{\baselineskip}
  \mbox{\small \Qcircuit @C=2em @R=1.4em {
  \lstick{}& \multigate{1}{ZX(\chi)} & \rstick{}\qw \\
  \lstick{}& \ghost{ZX(\chi)} & \rstick{}\qw
  }} \\
  (d)
  \end{minipage}
  \caption{\label{fig:NativeGates} Quantum gates native to various quantum computing architectures. Shown for each operation is the evolution of the qubit states (above) and the block circuit diagram with time going left to right (below). (a) The single-qubit rotation gate $R(\theta,\phi)$ creates superpositions according to two continuous parameters $\theta$ and $\phi$. (b) The $CZ$ operation on Rigetti systems performs a $Z$ rotation on a target qubit depending on a control qubit. (c) The $XX$ gate on IonQ systems and (d) the $ZX$ gate implemented in IBM superconducting systems both operate on two qubits with continuous parameter $\chi$ set to $\pi/4$.}
\end{figure}
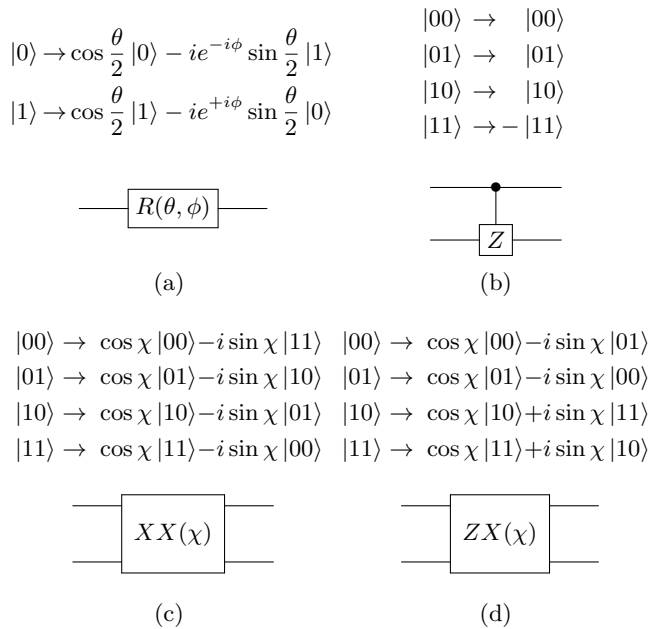

Ion trap quantum computers store qubits in the internal states of individual electrically charged atoms (ions). The ions are laser cooled and confined with electric fields from nearby electrodes, forming a 1D crystal in free space and in a vacuum environment, as shown in Fig. \ref{fig:QCs} \cite{wineland1998, Bruzewicz2019}.
Quantum gates are accomplished by poking the ions with focused laser beams that affect a qubit state-dependent force that modulates the Coulomb interaction between the ions and effectively ``wires" together any pair of qubits \cite{Monroe:2013}. The ion trap system we use, produced by IonQ, features ``rotation" gates $R(\theta,\phi)$ for single-qubit operations \cite{MikeAndIke} and Ising ($XX$) gates for two-qubit entanglement, as described in Fig. \ref{fig:NativeGates} \cite{Debnath2016}.

Superconducting quantum computer systems store qubits in Josephson junctions typically arrayed on a 2D lattice (see Fig. \ref{fig:QCs})  and cooled to very low temperature \cite{Kjaergaard2020}. Here, quantum gates are executed by running electrical currents through the device in a way that either changes the state of a single qubit or exploits the nonlinear interaction between adjoining qubits. The superconducting systems we use, produced by IBM and Rigetti, feature $R(\theta,\phi)$ rotation gates for single-qubit operations. The IBM systems use cross-resonance ($ZX$) gates for two-qubit entanglement \cite{IBMgates}, and Rigetti systems use controlled-Z ($CZ$) gates for two-qubit entanglement operations \cite{RigettiGates}.

\begin{figure}[h]
\includegraphics[width=\linewidth]{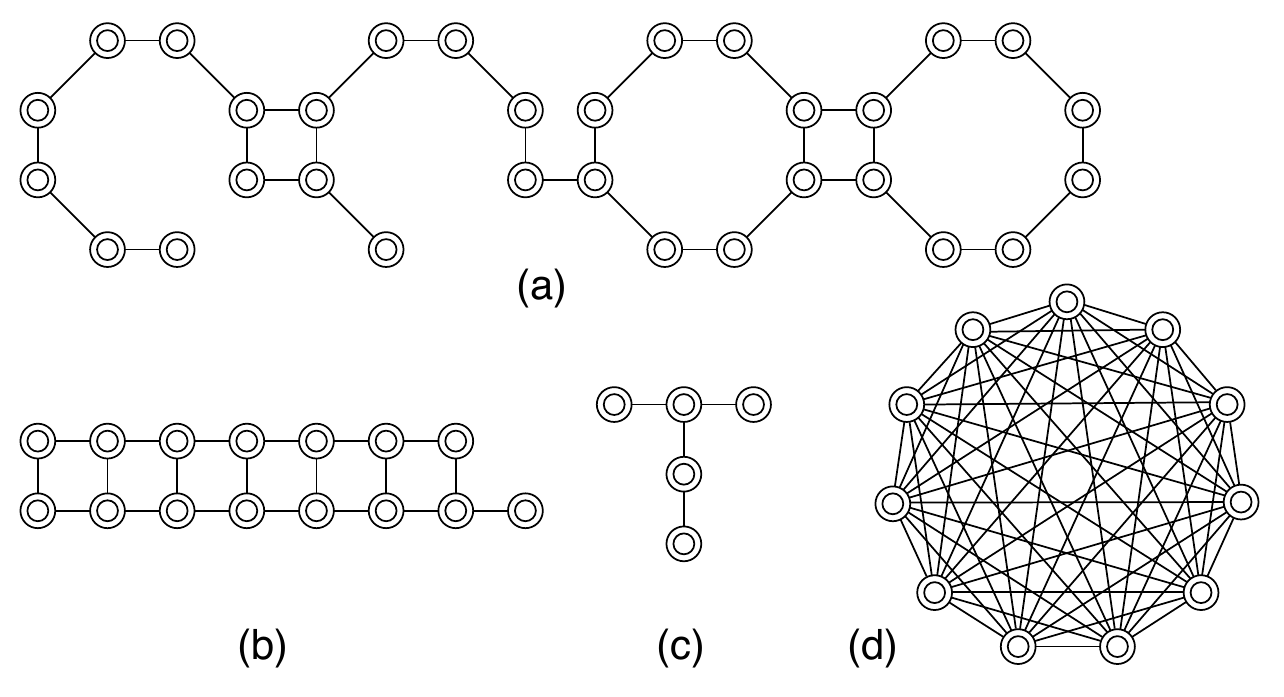}
\caption{Topology of qubit connectivity on the quantum computers used in this paper. Circles represent qubits, and lines represent the available two-qubit gates between qubits. (a) Rigetti Aspen-8 (31 qubits), (b) IBM-Melbourne (15 qubits), and (c) IBM-Vigo (5 qubits) superconducting quantum computers. (d) IonQ (11-qubit) ion trap quantum computer. \label{fig:graph}}
\end{figure}

The quantum computers we test each have a unique qubit layout or topology that forms a graph, whose vertices represent individual qubits and edges between vertices represent the ability to directly perform a two-qubit operation between the qubits. Connectivity characterizes how many distinct connections can be made between the vertices in the graph. When a two-qubit operation is performed between two qubits not directly connected, it must be decomposed into a series of SWAP (or other) operations involving intermediary qubits. The resultant increase in circuit complexity generally degrades performance for circuits that demand more connectivity than the system topology offers. Superconducting systems generally have fewer connections because of the two-dimensional wiring between nearest-neighbor Josephson junctions. On the other hand, ion trap quantum computers can be fully connected with direct gates available between any pair of qubits 
\cite{Linke:2017, Murali2019}. Fig. \ref{fig:graph} plots the connectivity graphs of the machines we used.

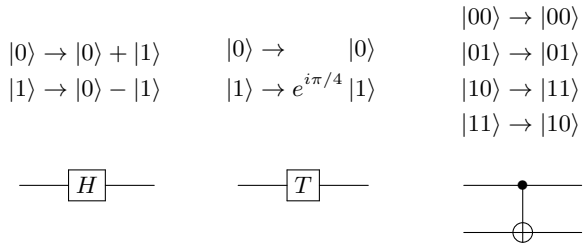
\begin{figure}[h] 
    \begin{minipage}[b]{0.32\linewidth}%
  \begin{eqnarray*}%
    \ket{0} &\rightarrow& \ket{0} + \ket{1}\\%
    \ket{1} &\rightarrow& \ket{0} - \ket{1}\\%
  \end{eqnarray*}%
  \mbox{\small \Qcircuit @C=2em @R=1.4em {\lstick{}& \gate{H} & \rstick{}\qw}}%
  \end{minipage}
\begin{minipage}[b]{0.32\linewidth}%
\begin{eqnarray*}%
\ket{0} &\rightarrow& \hspace{21pt}\ket{0}\\%
\ket{1} &\rightarrow& e^{i\pi/4} \ket{1}\\%
\end{eqnarray*}%
\mbox{\small \Qcircuit @C=2em @R=1.4em {\lstick{}& \gate{T} & \rstick{}\qw}}%
\end{minipage}
    \begin{minipage}[b]{0.33\linewidth}
    \begin{eqnarray*}%
        \ket{00} &\rightarrow& \ket{00}\\%
        \ket{01} &\rightarrow& \ket{01}\\%
        \ket{10} &\rightarrow& \ket{11}\\%
        \ket{11} &\rightarrow& \ket{10}%
    \end{eqnarray*}%
   \mbox{\small \Qcircuit @C=2em @R=1.4em {\lstick{}& \ctrl{1} & \rstick{}\qw \\ \lstick{}& \targ & \rstick{}\qw}}%
  \end{minipage}
  
\caption{\label{fig:StdGates} Standard universal quantum gates. The two single-qubit gates in this family are the Hadamard gate (left) and T gate (middle). The two-qubit CNOT gate (right) flips the second qubit if and only if the first qubit is in state $\ket{1}$.}
\end{figure}

Quantum algorithms and circuits are typically expressed in terms of a universal standard gate set, such as the single-qubit Hadamard and T gate plus the two-qubit Controlled-NOT (CNOT) gate, shown in Fig. \ref{fig:StdGates} \cite{MikeAndIke}. Running algorithms and circuits on a quantum computer generally requires a compilation step, where the standard gates are converted to native gates. For instance, the Hadamard gate is expressed as two rotations $H=R(\pi,0)R(\pi/2,\pi/2)$ and the CNOT gate is expressed as $CZ$, $XX$, or $ZX$ native gates with additional single qubit rotations \cite{Maslov2016}.  

Cloud quantum computing systems utilize transpilation, a heuristic process to simplify and optimize the circuits they run \cite{Chong2017}. For example, if multiple consecutive rotation operations are performed on a single qubit, the transpiler reduces them to a single rotation. When qubits are stored in disparate energy levels, perfoming a single-qubit phase gate \cite{MikeAndIke} can be replaced by a much simpler classical phase advance on the driving field for the next gate. Finally, operations with pure $\ket{0}$ or $\ket{1}$ inputs can be propagated classically, thus simplifying parts of the circuit.  There are many other compilation and transpilation techniques on cloud servers that are not obvious to the user, and this can make it difficult to extract performance metrics of the native components.

Quantum algorithms and circuits typically have three stages: initialization, coherent quantum gate operation, and measurement. Below, we perform a series of experiments on the cloud quantum computers to analyze the performance of each of these steps.

\section{State Preparation and Measurement}
We first examine the efficacy of state preparation and measurement (SPAM). We measure SPAM error by initializing a single qubit in either the $\ket{0}$ or $\ket{1}$ state, then immediately measuring the qubit. SPAM error is not expected to be symmetric, because the physical qubits occupy different energy levels and the detection process itself is not symmetric: one state generates a physical signal (an electrical current in the case of superconductors 
\cite{Kjaergaard2020} and photonic counts in the case of trapped ions \cite{Bruzewicz2019})
while the other state is a null signal. We observe that the superconducting systems have somewhat higher SPAM errors for the $\ket{1}$ state. We are unable to directly measure SPAM error for the $\ket{1}$ state on the IonQ computer, because for this trivial operation the transpiler simply complements the $\ket{0}$ measurement. However, this system is reported to have an average SPAM error (both states of all qubits) of $0.002$ \cite{Wright:2019}.

\begin{table}[h]
\begin{center}
\caption{Measured SPAM errors on the IonQ, IBM-Melbourne, and Rigetti Aspen-8 systems.}
\begin{tabular}[t]{l|cccc}
\hline
Quantum Comp. & SPAM & SPAM & SPAM & \\
 System & error $\ket{0}$ & error $\ket{1}$ & error avg.& $\#$ shots \\
\hline
IonQ & 0.0004 & 0.002* & 0.0012* & 3000 \\ 
IBM-Melbourne & 0.0253 & 0.0613 & 0.0433 & 3000 \\
Rigetti Aspen-8 & 0.047 & 0.175 & 0.111 & 3000 \\
\hline
\end{tabular}
\end{center}
  \vspace{-1em}
* inferred from previous work \cite{Wright:2019}.
  \label{SPAM}
\end{table}%

\vspace{-1em}

\section{Gates}
\begin{figure}[h] 
 \begin{minipage}[b]{1.0\linewidth}
\mbox{\small 
\Qcircuit @C=0.6em @R=1.1em{
    &  \lstick{\ket{0}} & \gate{H} & \gate{X} & \ctrl{1} & \gate{Y} & \ctrl{1} & \gate{X} & \ctrl{1} & \gate{Y} & \ctrl{1} & \gate{H} & \meter \\
    &  \lstick{\ket{0}} & \qw & \qw & \targ & \qw & \targ & \qw & \targ & \qw & \targ & \qw & \qw \gategroup{1}{4}{2}{7}{.7em}{-}
}}
\vspace{0.5\baselineskip}
  \end{minipage}
\caption{Circuit for testing two-qubit gate error accumulation in the IBM-Melbourne system. 
The boxed portion of the circuit is repeated a number of times. Note that only the upper qubit is measured. \label{fig:1qubitcirc}}
\end{figure}
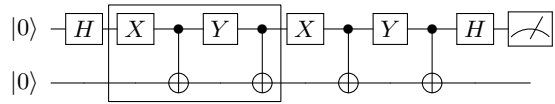

We characterize two-qubit gate quality by running simple circuits and measuring departures from the expected outputs. Fig. \ref{fig:1qubitcirc} shows an example circuit composed of a series of CNOT gates where we measure one of the output qubits. To bypass transpilation of the circuit, we surround the CNOT gates with $X \equiv R(\pi,0)$ and $Y \equiv R(\pi,\pi/2)$ rotation operations, and the entire circuit is surrounded by Hadamard gates. 
The effects of the single qubit gates are expected to be much smaller than the two-qubit CNOT operations.

For the IBM-Melbourne system, we observe a consistent decrease in performance with circuit depth, converging to a success rate of 0.5, which is expected from incoherent noise accumulation. The overall trend of circuit performance closely follows a Gaussian function with a characteristic $1/e$ degradation after approximately $28$ CNOT gates. 

\begin{figure}[h]
\includegraphics[width=\linewidth]{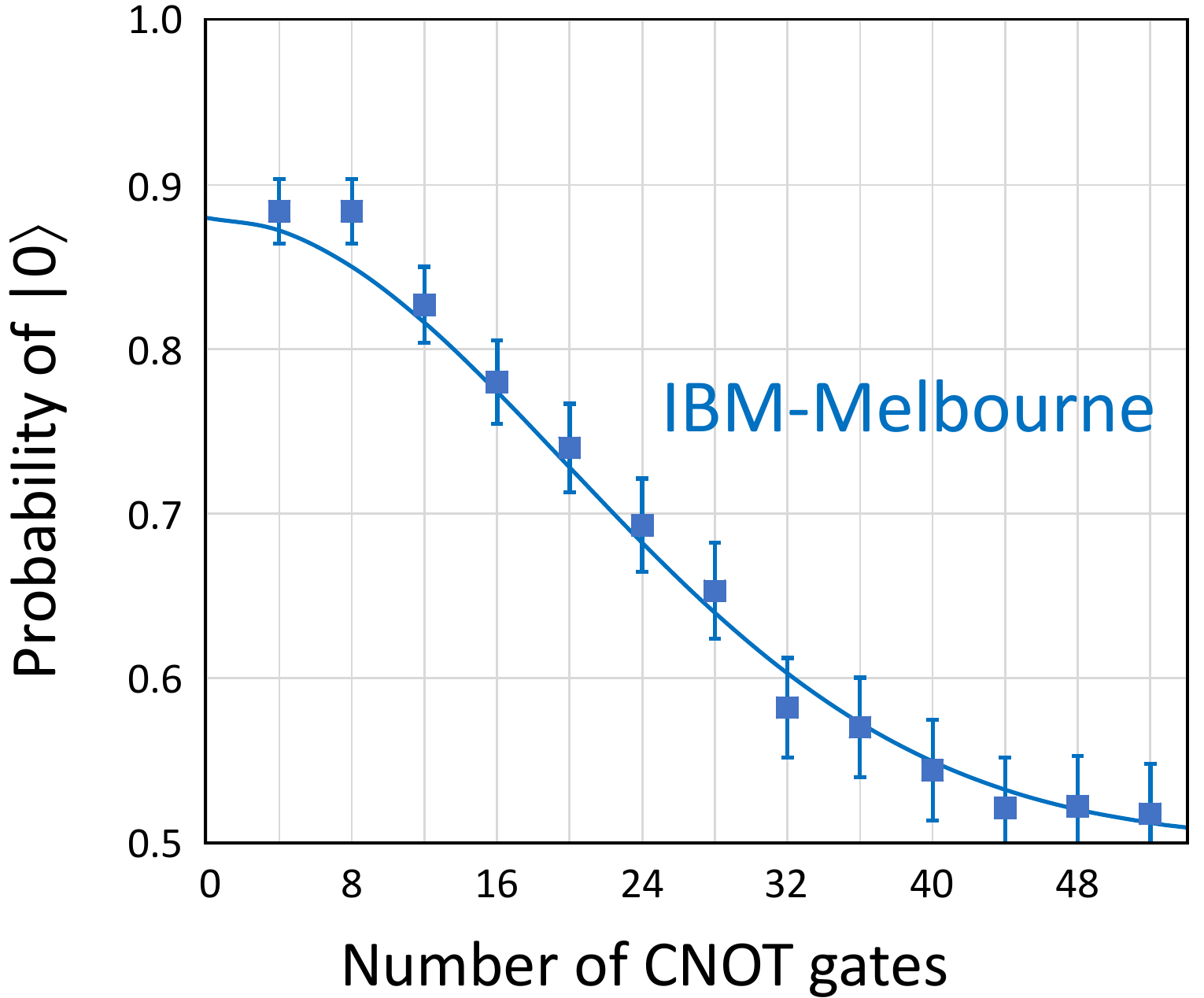}
\caption{Measured probability of the state $\ket{0}$ for the upper qubit in the circuit shown in Fig. \ref{fig:1qubitcirc} on the IBM-Melbourne system, as a function of the circuit depth. Solid line is a fit to a Gaussian function. Statistical error bars represent 95\% confidence intervals, using 1024 shots (samples) per run.
\label{fig:Rotation}}
\end{figure}

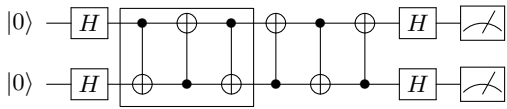
\begin{figure}[h] 
 \begin{minipage}[b]{1.0\linewidth}
\mbox{\small 
\Qcircuit @C=1em @R=1.1em{
    &  \lstick{\ket{0}} & \gate{H} & \ctrl{1} & \targ & \ctrl{1} & \targ & \ctrl{1} & \targ & \gate{H} & \meter \\
    &  \lstick{\ket{0}} & \gate{H} & \targ & \ctrl{-1} & \targ & \ctrl{-1} & \targ & \ctrl{-1} & \gate{H} & \meter \gategroup{1}{4}{2}{6}{1em}{-} 
}}
\vspace{0.5\baselineskip}
  \end{minipage}
\caption{Circuit for testing two-qubit gate error accumulation. The boxed portion of the circuit (a single SWAP) is repeated, to examine degradation in performance of larger circuits.\label{swaps}}
\end{figure}

On IonQ and Rigetti systems, transpilation of the circuit prevents a similar test of two-qubit gate fidelity. We instead run successive SWAP operations between two qubits initialized in the $\ket{0}$ state, as shown in Fig. \ref{swaps}. Again, we circumvent the transpiler reduction of the circuit by surrounding the sequence of SWAP gates with Hadamard gates. The error due to these additional gates (as well as the SPAM error) is much lower than the two-qubit gate error and also constant for each circuit, while the error due to the SWAP operations accumulates and is readily observed.

\begin{figure}[t]
\includegraphics[width=\linewidth]{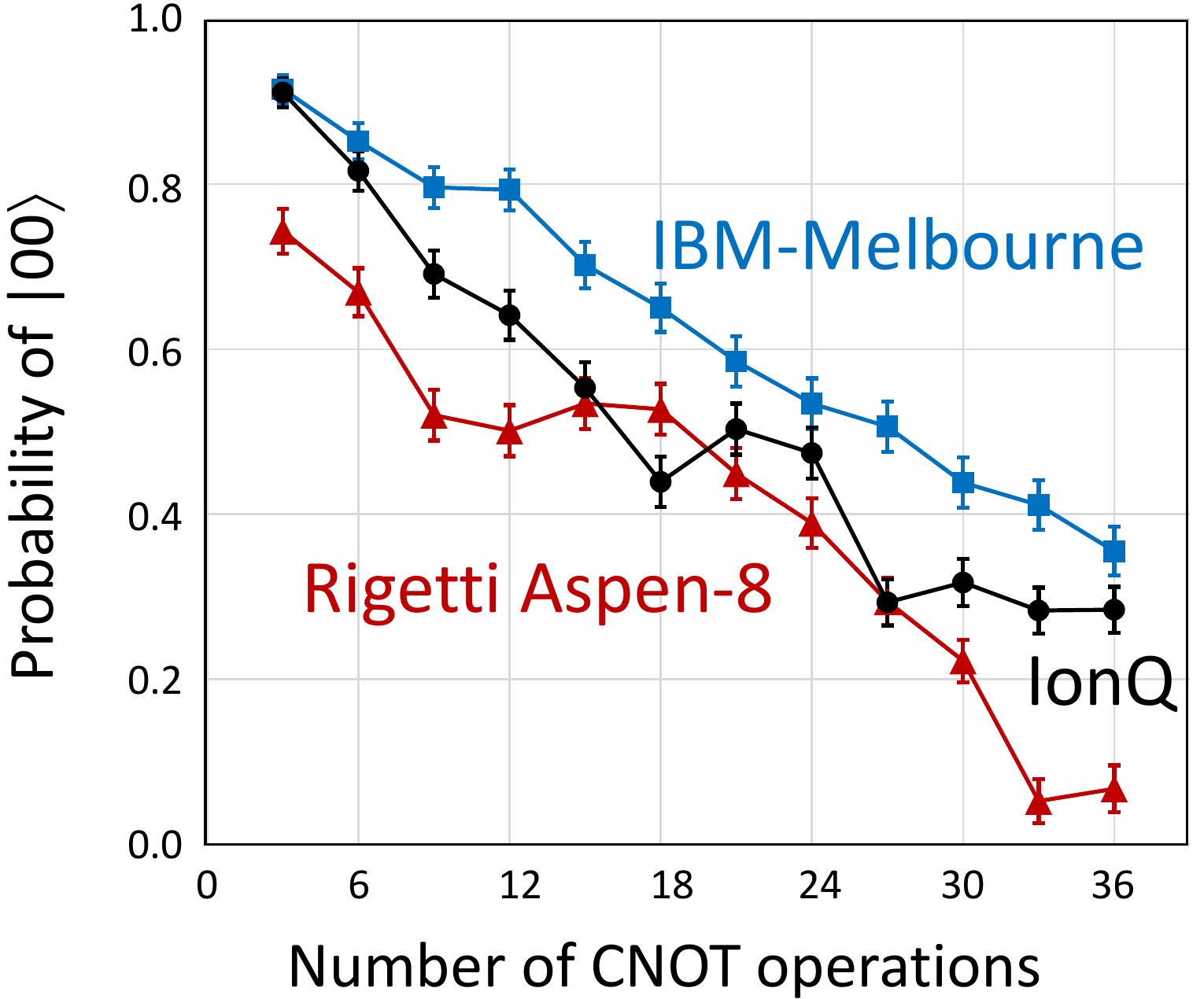}
\caption{Measured probability of the state $\ket{00}$ for the circuit shown in Fig. \ref{swaps} for the IonQ (black circles), IBM (blue squares), and Rigetti (red triangles) machines, as a function of circuit depth. Statistical error bars represent $95\%$ confidence intervals based on 1024 samples per point.}
\end{figure}

We observe a general decrease in circuit performance with respect to the number of SWAP operations performed. We expect the state probability to converge to 0.25, expected from incoherent noise accumulation. However, the observed decay does not follow an exponential law.  Moreover, the probability in the Rigetti Aspen-8 system dropped to below 0.05 for the 11 and 12 SWAP operations, possibly indicating the presence of coherent or systematic errors.

Extrapolation of success rates to 0 SWAP gates gives a rough estimate of constant (depth-independent) noise levels on each system. We use a linear regression model on the first 4 data points for each system, to approximate the relationship between performance and circuit depth for smaller circuits. For IonQ, this model yields an intercept of 0.999, with each set of three CNOT gates decreasing success probability by 0.0935. This intercept is not far from the SPAM error observed for IonQ in Table I. For IBM-Melbourne, this model yields an intercept of 0.945, with each set decreasing success probability by 0.0422. Again, this is roughly consistent with the experimentally determined SPAM error. For Rigetti Aspen-8, this model yields an intercept of 0.827, with each set decreasing success probability by 0.0875. 

Rigetti Aspen-8 success rates are inconsistent between runs, with probabilities sometimes deviating from the average by $0.2$-$0.3$. This may be attributed to different qubit assignment between runs, calibration/degradation of the entire system, or other unknown factors. 


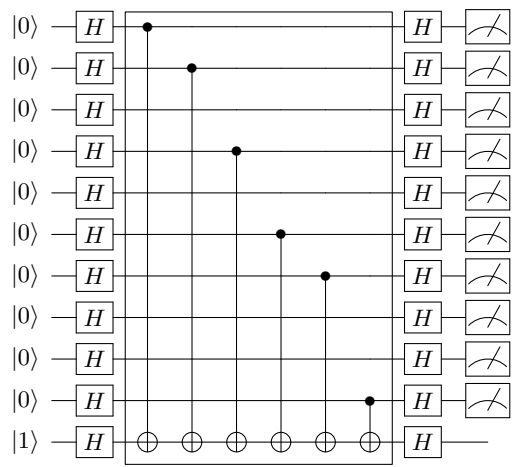
\begin{figure}[b] 
 \begin{minipage}[b]{1.0\linewidth}
\mbox{\small 

\Qcircuit @C=1.0em @R=0.3em @!R {
&  & \lstick{\ket{0}} & \gate{H}&\ctrl{10}&\qw     &\qw     &\qw     &\qw     &\qw     &\gate{H}&\meter\\
&  & \lstick{\ket{0}} &\gate{H}&\qw      &\ctrl{9}&\qw     &\qw     &\qw     &\qw     &\gate{H}&\meter\\
&  & \lstick{\ket{0}} &\gate{H}&\qw      &\qw     &\qw     &\qw     &\qw     &\qw     &\gate{H}&\meter\\
&  & \lstick{\ket{0}} &\gate{H}&\qw      &\qw     &\ctrl{7}&\qw     &\qw     &\qw     &\gate{H}&\meter\\
&  & \lstick{\ket{0}} &\gate{H}&\qw      &\qw     &\qw     &\qw     &\qw     &\qw     &\gate{H}&\meter\\
&  & \lstick{\ket{0}} &\gate{H}&\qw      &\qw     &\qw     &\ctrl{5}&\qw     &\qw     &\gate{H}&\meter\\
&  & \lstick{\ket{0}} &\gate{H}&\qw      &\qw     &\qw     &\qw     &\ctrl{4}&\qw     &\gate{H}&\meter\\
&  & \lstick{\ket{0}} &\gate{H}&\qw      &\qw     &\qw     &\qw     &\qw     &\qw     &\gate{H}&\meter\\
&  & \lstick{\ket{0}} &\gate{H}&\qw      &\qw     &\qw     &\qw     &\qw     &\qw     &\gate{H}&\meter\\
&  & \lstick{\ket{0}} &\gate{H}&\qw      &\qw     &\qw     &\qw     &\qw     &\ctrl{1}&\gate{H}&\meter\\
&  & \lstick{\ket{1}} &\gate{H}&\targ    &\targ   &\targ   &\targ   &\targ   &\targ   &\gate{H} &\qw   
 \gategroup{1}{5}{11}{10}{1em}{-}
}}
    \vspace{0.5\baselineskip}
   \end{minipage}
\caption
{10-qubit Bernstein-Vazirani circuit with example of a 10-bit hidden string $\textbf{a}= 1101011001$ encoded in the (boxed) oracle using CNOT gates. For each qubit, the presence of a CNOT gate to the 11th ancillary qubit (bottom) indicates the corresponding bit of the hidden string is in the state '1'. \label{fig:BVcircuit}}
\end{figure}

\begin{figure*}[ht]
\includegraphics[width=0.9\linewidth]{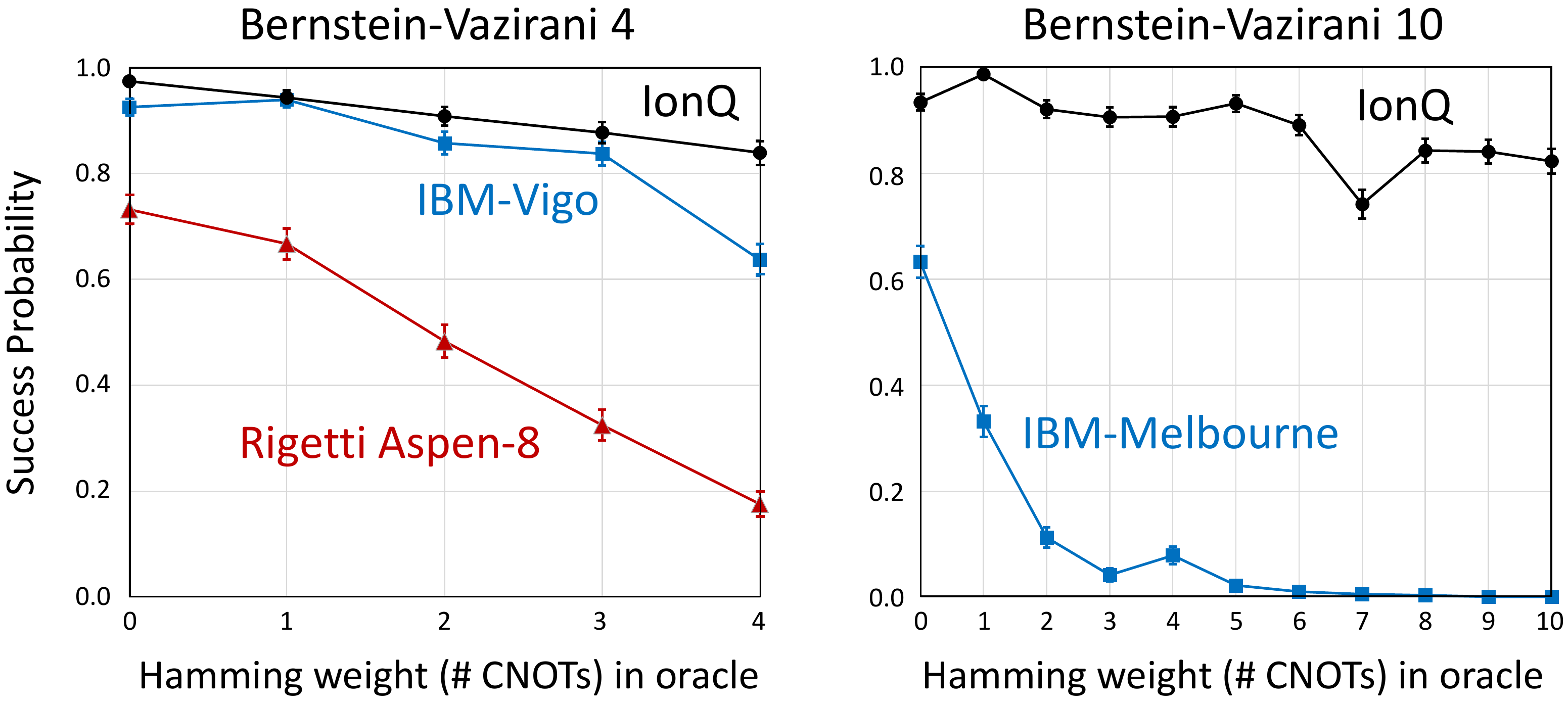}
\caption{Measured success probability for 4-qubit and 10-qubit Bernstein-Vazirani algorithm as a function of the Hamming weight in the oracle, or the number of oracle CNOT gates to the ancillary qubit. Data is presented from IonQ (black circles), IBM-Vigo (blue squares, left) IBM-Melbourne (blue squares, right), and Rigetti Aspen-8 (red triangles) machines. (The Rigetti system did not produce satisfactory output for the 10 qubit algorithm).  Statistical error bars represent 95\% confidence intervals based on 1024 samples per point. \label{fig:BVdata}}
\end{figure*}

\section{Scaling Circuit Example}
We investigate larger-scale performance of the quantum computers by implementing the Bernstein-Vazirani (BV) algorithm \cite{BV}.
The BV oracle encodes a hidden $n$-bit string $\textbf{a}$, and when queried with a $n$-bit string $\textbf{x}$, returns the (single bit) dot product $\textbf{a}$ $\cdot$ $\textbf{x}$. Classically, the best method for determining $\textbf{a}$ is to query the oracle $n$ times, once for each bit of $\textbf{a}$. A quantum computer can return $\textbf{a}$ with just a single query, by presenting a superposition of all possible strings $\textbf{x}$. We implement the BV oracle by using specific patterns of CNOT gates to represent given hidden strings $\textbf{a}$ (Fig. \ref{fig:BVcircuit}). We run the algorithm with both 4- and 10-bit oracles. 

As the hidden bit string $\textbf{a}$ increases in Hamming weight, more CNOT gates are performed with an ancillary qubit. This will degrade performance, due to both the number of additional CNOT gates and also the greater number of connections required to the ancillary qubit. We observe both effects, as shown in Fig. \ref{fig:BVdata}. 
All of the systems show a general negative trend with the Hamming weight of the oracle. 

Interestingly, in the BV-4 algorithm, we observe a significant drop in performance on the IBM-Vigo system at a Hamming weight of 4 (requiring 4 CNOT connections to the ancilla). This is consistent with the limited connectivity of the machine, which has at most 3 connections from a given qubit, as shown in Fig. \ref{fig:graph}c. Even optimal implementation of BV on this system would still only allow for three direct CNOT connections. 

On BV-10, the IBM-Melbourne system shows a steep drop in performance beyond a Hamming weight of 1-2. 
This may be attributed to crosstalk errors, where noise from a qubit can influence adjacent qubits, resulting in lower overall success rate for any number of CNOT operations.
Additionally, IBM-Melbourne's nearest-neighbor connectivity (Fig. \ref{fig:graph}b) only allows for a maximum of 3 direct connections to a single qubit, requiring additional (noisy) SWAP operations. The IonQ system shows a modest drop in performance with Hamming weight, but because of its full connectivity, this is consistent with simple accumulated gate errors.

It should be noted that the best classical success rate for successfully determining the hidden n-bit string is $1/2^{n-1}$; because only one query of the BV oracle is allowed, the classical solution determines one bit of the string and guesses all the others. Thus, for BV-10, even a 1\% success rate is better than the optimal classical solution.

\section{Outlook}

While comparing circuit and algorithmic performance on these cloud quantum computer systems is straightforward, it is difficult to extract component performance or predict performance on other complex circuits. Our measurements show strong variations in circuit performance, possibly due to compiler/transpiler processes native to each system. For example, when comparing the two-qubit SWAP sequence of Fig. \ref{swaps} vs. the BV-10 circuit in Fig. \ref{fig:BVdata}, the inferred two-qubit gate errors in each system differ significantly. This could be a result of each system's compiler or transpiler simplifying circuits in a different way. However, BV performance also depends critically on connectivity and crosstalk errors, which vary strongly between the systems.

A drawback in our analysis of these quantum computer cloud systems is a lack of information on aspects such as qubit assignment, compiler/transpiler methods, component drift rate, and time since last calibration.
With significant time delay between submitting a circuit to the cloud and receiving a result, we cannot pinpoint the times at which calibrations occur, and how these errors influence our findings. We observe significant fluctuations in circuit performance between runs on all systems, and we suspect these systematic errors overwhelm any statistical errors. In the future, it will be interesting to run the same circuit over a wider range of time, which, along with access to detailed information on component performance, could control for such errors. 

\section{Acknowledgements}
We acknowledge useful discussions with B. Blinov (University of Washington), Y. Nam (IonQ), J. Gambetta (IBM), and D. Zhu (University of Maryland). C. Monroe is co-founder and Chief Scientist at IonQ, Inc. 



%

\end{document}